# The stability of MHD Taylor-Couette flow

RÜDIGER, G.
Astrophysikalisches Institut Potsdam, An der Sternwarte 16, D-14482 Potsdam, GERMANY
e-mail address: gruediger@aip.de

**Abstract:** The linear marginal instability of an MHD Taylor-Couette flow of infinite vertical extension is considered. For hydrodynamically unstable flows the minimum Reynolds number exists even without a magnetic field, but there are also solutions with smaller Reynolds numbers for certain (weak) magnetic fields. The magnetic field can, therefore, destabilize the rotating flow. This instability, however, can only exist for not too small magnetic Prandtl numbers. Also the situation for nonaxisymmetric disturbances is considered. Without magnetic fields such modes possess higher Reynolds numbers than the axisymmetric modes. The situation is changed in the MHD regime. For modest magnetic fields the excitation of modes with *m*=1 is easier than the excitation of modes with *m*=0 if critical magnetic amplitudes are exceeded.

## 1. Introduction

The longstanding problem of the generation of turbulence in various hydrodynamically stable situations has found a solution in recent years with the so-called magnetorotational instability (MRI), in which the presence of a magnetic field has a destabilizing effect on a differentially rotating flow, provided that the angular velocity decreases outwards with the radius.

However, the MRI has never been observed in the laboratory (see Ji et al. 2001). Moreover, Chandrasekhar (1961) already suggested the existence of the MRI for ideal Taylor-Couette flow, but his results for non-ideal fluids for small gaps and within the small magnetic Prandtl number approximation demonstrated the absence of the MRI for hydrodynamically unstable flow. Recently, Goodman and Ji (2002) claimed that this absence of MRI was due to the use of the small magnetic Prandtl number limit. The magnetic Prandtl number is really very small under laboratory conditions ($\sim 10^{-5}$ and smaller). Obviously, a proper understanding of this phenomenon is very important for possible future experiments, including *Taylor-Couette flow dynamo experiments*.

The dependence of a real Taylor-Couette flow on the magnetic Prandtl number is investigated also for nonaxisymmetric modes. For viscous flows, in the absence of any transverse pressure gradient, the most general form of $\Omega$ allowed is $\Omega(R) = a + b/R^2$ where $a$ and $b$ are two constants related to the angular velocities $\Omega_{in}$ and $\Omega_{out}$ with which the inner and the outer cylinders are rotating. If $R_{in}$ and $R_{out}$ are the radii of the two cylinders then

$$a = \Omega_{in} \frac{\hat{\mu} - \hat{\eta}^2}{1 - \hat{\eta}^2} \qquad \text{and} \qquad b = \Omega_{in} R_{in}^2 \frac{1 - \hat{\mu}}{1 - \hat{\eta}^2} \qquad (1)$$

with $\hat{\mu} = \Omega_{out}/\Omega_{in}$ and $\hat{\eta} = R_{in}/R_{out}$. Rotation laws are hydrodynamically stable for $\hat{\mu} > \hat{\eta}^2$. Taylor-Couette flows with outer cylinders at rest $\hat{\mu} = 0$ are thus never stable.

Only one sort of containers is here considered with one and the same gap geometry, i.e. $\hat{\eta}=$ 0.5. Then the flow between the cylinders is hydrodynamically unstable between $\hat{\mu}=0$ and $\hat{\mu}=0.25$. We shall work with both the hydrodynamically unstable container with $\hat{\mu}=0$ and with the hydrodynamically stable container with $\hat{\mu}=0.33$. We are interested in the stability of the internal rotation law between the cylinders.

By analyzing the disturbances into normal modes, the solutions of the linearized MHD equations are of the form

$$(\boldsymbol{u'}, \boldsymbol{B}) = (\boldsymbol{u}(R), \boldsymbol{B}(R))\, e^{i(m\varphi + kz - \omega t)}. \qquad (2)$$

Only marginal stability will be considered. The derivation of the equations describing this situation is given in Rüdiger & Shalybkov (2002). The Reynolds number Re and the Hartmann number Ha are defined as

$$\mathrm{Re} = \frac{\Omega_{in} R_{in}(R_{out}-R_{in})}{\nu}, \qquad \mathrm{Ha} = B_0 \sqrt{\frac{R_{in}(R_{out}-R_{in})}{\mu_0 \rho \nu \eta}}. \qquad (3)$$

For given Hartmann number and magnetic Prandtl number, Pm= $\nu/\eta$ ($\nu$ is the kinematic viscosity, $\eta$ is the magnetic diffusivity) we shall derive in a linear theory the critical Reynolds number of the rotation of the inner cylinder for various mode numbers *m*.

An appropriate set of ten boundary conditions is needed to solve the system. No-slip conditions for the velocity on the walls are used, i.e. $u_R=0$, $u_\phi=0$, $du_R/dR=0$. The magnetic boundary conditions depend on the electrical properties of the walls. The transverse currents and perpendicular component of magnetic field should vanish on conducting walls, hence $u_R = 0$, $u_\phi = 0$, $du_R/dR = 0$. The above boundary conditions are valid for $R=R_{in}$ and for $R=R_{out}$.

## 2. The axisymmetric modes

In Figure 1 an outer cylinder at rest is considered for Pm=1 (left). For vanishing magnetic field and for $\eta=0.5$ the exact Reynolds number for this case is 68.2. For increasing magnetic field the Reynolds number is reduced so that the excitation of the Taylor vortices becomes easier than without magnetic field. The minimum Reynolds number $\mathrm{Re}_{\mathrm{crit}}$ of about 63 for Pm=1 is reached for $\mathrm{Ha}_{\mathrm{crit}} \cong 4...5$. This magnetic-induced subcritical excitation of Taylor vortices is due to the MRI. For stronger magnetic fields the instability is suppressed by the magnetic field so that the Reynolds number starts to grow. In Figure 1 the same container is also considered but for a small magnetic Prandtl number of $10^{-5}$ (right). The minimum, which is well pronounced for the Pm=1 case, completely disappears. A suppression of the instability by the magnetic field can be observed. Obviously, the MRI does not work efficiently in the limit of small magnetic Prandtl numbers, i.e. for too low electrical conductivity.

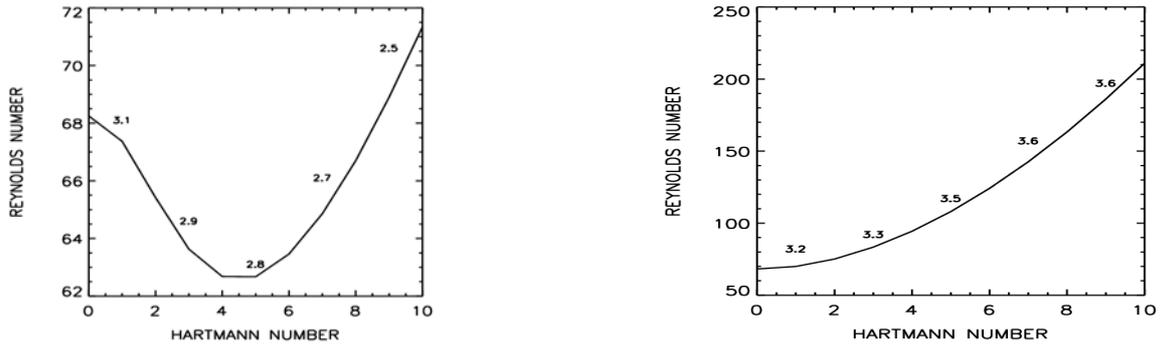

Fig. 1: The stability line for Taylor-Couette flow with outer cylinder at rest for Pm =1 (left) and Pm=$10^{-5}$ (right). The flow is unstable above the line. There is instability even without magnetic fields but for Pm=1 its excitation is easier with magnetic fields with Ha $\cong$ 4.5. The lines are marked with those wave numbers for which the Reynolds numbers are minimum (see Rüdiger & Shalybkov 2002)

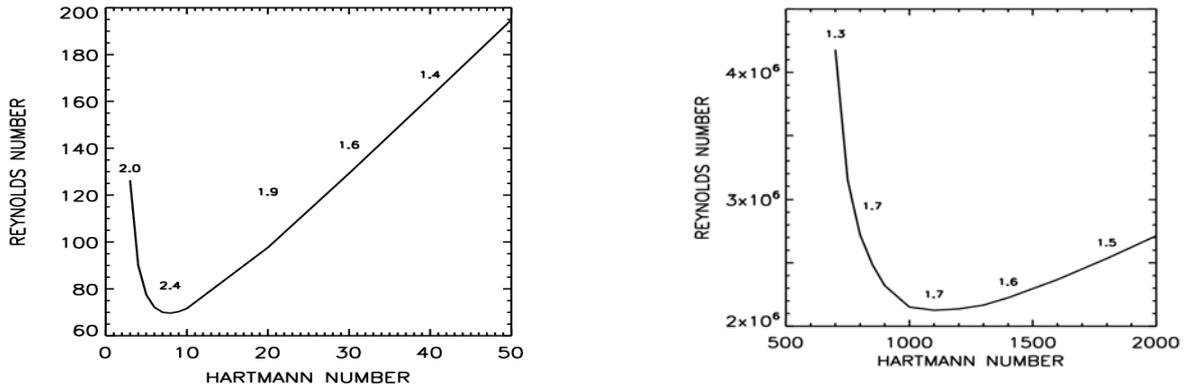

Fig. 2: The stability line for Pm =1 (left) and Pm=$10^{-5}$ (right). The outer cylinder rotates with 33% of the rotation rate of the inner cylinder so that, from the Rayleigh criterion, the hydrodynamic instability for Ha =0 disappears. The minimum Reynolds number is almost the same as on the LHS of Fig. 1

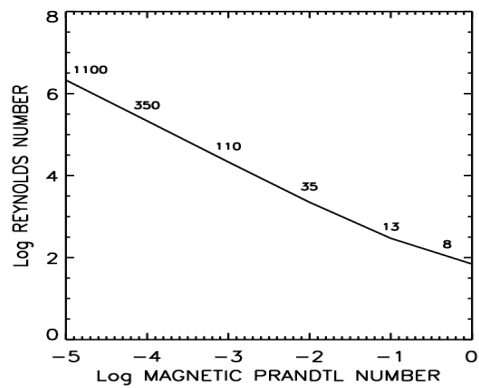

Fig. 3: The main results for $\hat{\eta}$ =0.5 and $\hat{\mu}$ =0.33: The critical Reynolds numbers for given magnetic Prandtl numbers marked with those Hartmann numbers where the Reynolds number is minimum

Another situation occurs if the outer cylinder rotates so fast that the rotation law does not longer fulfill the Rayleigh criterion, and a solution for Ha=0 cannot exist. Then the nonmagnetic eigenvalue along the vertical axis moves to infinity and we should always have a minimum. This is the basic situation in astrophysical applications such as accretion disks with a Kepler rotation law. The main question is whether the critical Reynolds number and the critical Hartmann number can be realized experimentally. The Figure 3 presents the results for $\hat{\mu} = 0.33$ and for Pm =1 (left) and Pm =$10^{-5}$ (right).

There are always minima of the characteristic Reynolds numbers for certain Hartmann numbers. The minima and the critical Hartmann numbers increase for decreasing magnetic Prandtl number. The critical Reynolds numbers together with the critical Hartmann numbers are plotted in Figure 3. Note that here the slope of the curve is much steeper than in the case of a container with finite vertical extension (Rüdiger & Zhang 2001, Willis & Barenghi 2002).

For small magnetic Prandtl numbers we find rather simple relations. With the magnetic Reynolds number Rm= Pm Re and the Lundquist number Ha* = √Pm Ha it follows Rm ≅ 21 and Ha* ≅ 3.5 (Rüdiger & Shalybkov 2002). For small Pm both quantities do not depend on the microscopic viscosity. Both the minimum magnetic Reynolds number and the corresponding characteristic Lundquist number are thus independent of the value of the kinematic microscopic viscosity. Hence, $B$= 1240 Gauss / (($R_{in}$/10 cm)) for small magnetic Prandtl number, and for the frequency of the inner cylinder we obtain $f$= 33 Hz/ ($R_{in}$/10 cm)$^2$. Note that the Hartmann number is maximum for the experiment with $\hat{\eta}$=0.5.

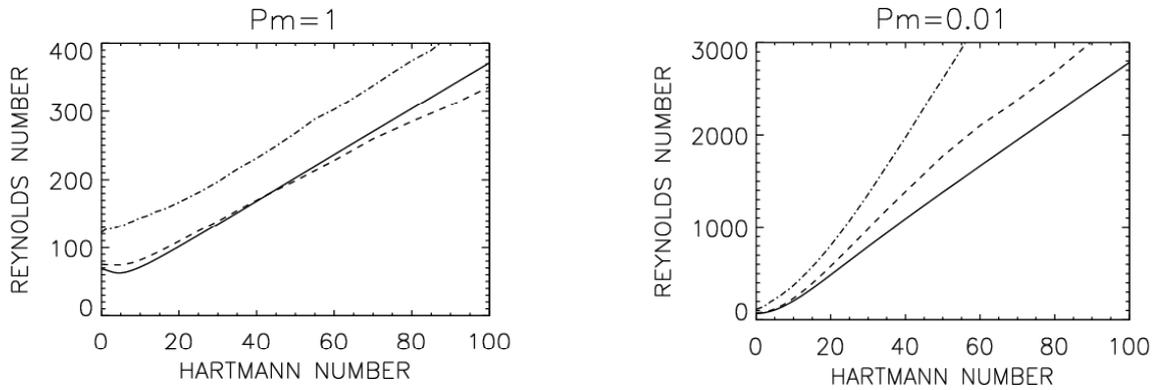

Fig. 4: Resting outer cylinder: Stability lines for axisymmetric (*m*=0, solid lines) and nonaxisymmetric instability modes (*m*=1 (dashed lines), *m*=2 (dashed-dotted lines)). Results are given for Pm=1 and Pm=0.01. Note that for Pm=1 for certain magnetic fields the nonaxisymmetric modes with *m*=1 possess the lowest Reynolds numbers

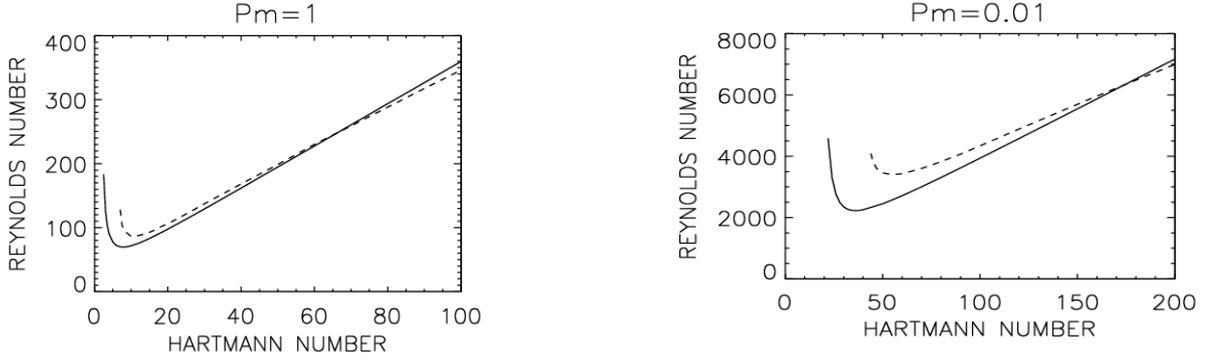

Fig. 5: The same as in Figure 4 but for rotating outer cylinder ($\hat{\mu}$=0.33), there is no hydrodynamical instability. Again for Hartmann numbers exceeding those if the minima, modes with *m*=1 possess the lowest eigenvalues

## 3. The nonaxisymmetric modes

The homogeneous set of equations with the boundary conditions for conducting walls determine the eigenvalue problem of the form

$$\mathcal{L}(k, m, \text{Re}, \mathcal{R}(\omega)) = 0 \tag{4}$$

for given Ha and Pm. The real part of ω, $\mathcal{R}(\omega)$, describes a drift of the pattern along the azimuth which only exists for nonaxisymmetric flows. $\mathcal{L}$ is a complex quantity, both its real part and its imaginary part must vanish for the critical Reynolds number. The latter is minimized by choice of the wavenumber *k*. $\mathcal{R}(\omega)$ is the second quantity which is fixed by the eigenequation (Shalybkov et al. 2002).

We start with the results for containers with resting outer cylinders (Figure 4). For Ha=*m*=0 the known critical Reynolds number Re=68 is reproduced. For *m*>0 the critical Reynolds numbers exceed the values for *m*=0. Without magnetic field the instability yields axisymmetric rolls. The critical Reynolds number for *m*=1 is 73 and for *m*=2 it is 101 (Roberts 1965). With magnetic fields (Ha>0) the magnetic Prandtl number comes into the game. Results for Pm=1 and 0.01 are presented in Figure 4. For Pm ≅1 the electrical conductivity is so high that the magnetorotational instability for Ha ≅5 produces a characteristic minimum of the critical Reynolds number but for stronger magnetic fields the suppressing action of the magnetic field starts to dominate. *In contrast to the expectations, however, for* Pm=1 *for increasing Hartmann number the mode with m=1 becomes more and more dominant*. This is a new and interesting result: The linear instability of the Taylor-Couette flow without magnetic field is formed by axisymmetric rolls but the magnetic field favours the excitation of bisymmetric spirals. For Ha ≥ 50 the instability sets in in form of a drifting pattern with maximum and minimum separated by 180º. As can also be seen in Figure 4 for small magnetic Prandtl number (here Pm=0.01) the axisymmetric pattern with *m*=0 again starts to dominate with the lowest critical Reynolds number. The modes with *m*=2, which we have also computed, do never possess the lowest Reynolds numbers, they are not important for the discussion of the pattern of the instability. What we have found is that in contrast to the hydrodynamic case there are combinations where the nonaxisymmetric mode with *m*=1 has a lower Reynolds number than the axisymmetric mode with *m*=0. This is one of the most surprising structure-forming consequences of the inclusion of magnetic fields to the Taylor-Couette flow experiment found first in astrophysical simulations.

If the outer cylinder rotates with $\hat{\mu}$ =0.33, then the linear instability without magnetic field disappears and the critical Reynolds number for Ha =0 moves to infinity. However, for finite Hartmann number (again of order 10) the instability survives with almost the same Reynolds numbers. The consequence is the occurrence of typical minima in the stability diagrams. The minima also occur for the nonaxisymmetric solutions with $m$=1. For Pm ≤ 1 we always find intersections between the lines for $m$=0 and $m$=1. Again there are critical Hartmann numbers at which the ring geometry ($m$=0) of the excited flow and field pattern changes to a nonaxisymmetric spiral geometry with $m$=1 (Figure 5).

Hence, also in experiments with rotating outer cylinder the magnetic field is able to produce nonaxisymmetric structures. After the Cowling theorem which requires the existence of nonaxisymmetric magnetic modes for the complete existence of a dynamo, a selfexcited dynamo might thus exist, but so far we have only results for magnetic Prandtl numbers between 0.01 and 1. The magnetic Prandtl number for experiments with liquid metals like sodium or gallium with Pm of order $10^{-(5....6)}$ are still smaller than the considered values (see Figure 6).

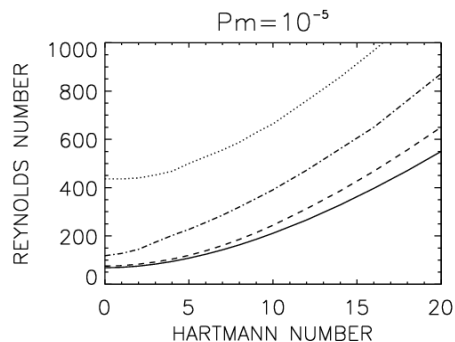

Fig. 6: The same as in Figure 4 but for the small Prandtl number $10^{-5}$ (for liquid sodium). Note that for such small values the axisymmetric mode for $m$=0 always seems to dominate. The results are preliminary

## 4. References

Chandrasekhar, S. 1961, Hydrodynamic and Hydromagnetic Stability (Clarendon, Oxford)
Goodman, J. & Ji, H. 2002, JFM, 462, 365
Ji, H., Goodman, J. & Kageyama, A. 2001, MNRAS, 325, L1
Roberts, P. 1965, Proc. Roy. Soc. London A, 283, 550
Rüdiger, G. & Zhang, Y. 2001, A&A, 378, 302
Rüdiger, G. & Shalybkov, D.A. 2002, Phys. Rev. E (in press)
Shalybkov, D.A., Rüdiger, G. & Schultz, M. 2002, A&A (subm.)
Willis, A.P., & Barenghi, C.F. 2002, A&A, 388, 688